\documentclass[pra,showpacs,amsfonts,amssymb,amsmath,eqsecnum,tightenlines,nofootinbib,showkeys]{revtex4}

\newcommand{\vs}{\vspace{0.25cm}}

\newtheorem{theorem}{Theorem}
\newtheorem{itlemma}{Lemma}[section]
\newtheorem{itproposition}[itlemma]{Proposition}
\newtheorem{itcorollary}[itlemma]{Corollary}
\newtheorem{itremark}[itlemma]{Remark}
\newtheorem{itremarks}[itlemma]{Remarks}
\newtheorem{itdefinition}[itlemma]{Definition}
\newtheorem{itexample}[itlemma]{Example}

\newenvironment{lemma}{\begin{itlemma}\rm}{\end{itlemma}} 
\newenvironment{remark}{\begin{itremark}\rm}{\end{itremark}} 
\newenvironment{remarks}{\begin{itremarks} \rm}{\end{itremarks}}
\newenvironment{corollary}{\begin{itcorollary}\rm}{\end{itcorollary}}
\newenvironment{proposition}{\begin{itproposition}\rm}{\end{itproposition}}
\newenvironment{definition}{\begin{itdefinition}\rm}{\end{itdefinition}}
\newenvironment{example}{\begin{itexample}\rm}{\end{itexample}}
\newenvironment{fact}{\noindent {\em Fact}. \ \ }{\hfill \medskip}
\newenvironment{proof}{\noindent {\em Proof}.\ \ }{\hspace*{\fill}$\Box$\medskip}
\newenvironment{claim}{\noindent {\em Claim}. \ \ }{\hfill \medskip}
\newcommand{\be}[1]{\begin{equation}\label{#1}}
\newcommand{\ee}{\end{equation}}
\newcommand{\bl}[1]{\begin{lemma}\label{#1}}
\newcommand{\br}[1]{\begin{remark}\label{#1}}
\newcommand{\brs}[1]{\begin{remarks}\label{#1}}
\newcommand{\bt}[1]{\begin{theorem}\label{#1}}
\newcommand{\bd}[1]{\begin{definition}\label{#1}}
\newcommand{\bp}[1]{\begin{proposition}\label{#1}}
\newcommand{\bc}[1]{\begin{corollary}\label{#1}}
\newcommand{\bfact}[1]{\begin{fact}\label{#1}}
\newcommand{\bex}[1]{\begin{example}\label{#1}}
\newcommand{\ec}{\end{corollary}}
\newcommand{\efact}{\end{fact}}
\newcommand{\eex}{\end{example}}
\newcommand{\el}{\end{lemma}}
\newcommand{\er}{\end{remark}}
\newcommand{\ers}{\end{remarks}}
\newcommand{\et}{\end{theorem}}
\newcommand{\ed}{\end{definition}}
\newcommand{\ep}{\end{proposition}}
\newcommand{\epr}{\end{proof}}
\newcommand{\bpr}{\begin{proof}}
\newcommand{\bcl}{\begin{claim}}
\newcommand{\ecl}{\end{claim}}

\newcommand{\bi}{\begin{itemize}}
\newcommand{\ei}{\end{itemize}}
\newcommand{\ben}{\begin{enumerate}}
\newcommand{\een}{\end{enumerate}}
\newcommand{\go}[1]

\begin{document}

\title{Decompositions of unitary evolutions and entanglement
dynamics of bipartite quantum systems}
\thanks{Work supported by NSF Career Grant, ECS0237925}

\author{Domenico D'Alessandro}
\email{daless@iastate.edu}
\author{Raffaele Romano}
\email{rromano@iastate.edu} \affiliation{Department of Mathematics,
Iowa State University, Ames IA 50011, USA}

\begin{abstract}

We describe a decomposition of the Lie group of unitary evolutions
for a bipartite quantum system of arbitrary dimensions. The
decomposition is based on a recursive procedure which systematically
uses the Cartan classification of the symmetric spaces of the Lie
group $SO(n)$. The resulting factorization of unitary evolutions
clearly displays  the local and entangling character of each factor.

\end{abstract}

\pacs{03.65.-w, 03.67.Mn, 02.20.Hj}

\keywords{decompositions of Lie groups, entanglement dynamics,
quantum information}

\maketitle


\section{Introduction}

Decompositions of a Lie group $G$ are methods to factorize every
element $X \in G$ as \be{DefLGD} X  = X_1 X_2 \cdots X_m, \ee where
the factors $X_1, \ldots, X_m$ belong to one dimensional subgroups.
Decompositions of unitary evolutions in simpler terms are of
interest in quantum control and information theory for at least
three reasons. They allow to simplify the task of controlling the
evolution of a quantum system to a target into a sequence of simpler
subtasks, consisting of control problems to more easily reachable
targets (see e.g.~\cite{MikoMCSS,RamaK,Schirm}). They allow the
analysis of several features of quantum dynamics such as
entanglement generation, time optimality and parameter
identification (see e.g.~\cite{BBO1,Zhang,KG,ConfraId}). They give
methods to produce unitary evolutions in a laboratory by combining a
sequence of readily reproducible evolutions. In particular, in
quantum information theory, a decomposition can be seen as a method
to generate a quantum logic operation from a sequence of elementary
operations~\cite{Lloyd,KG}.

In many cases, decompositions of the unitary group $U(n)$ correspond
to vector space decompositions of the corresponding Lie algebra
${\mathfrak u}(n)$ and each element $H \in i{\mathfrak u}(n)$
represents the Hamiltonian of a possible evolution. In the analysis
of multipartite quantum systems, it is useful to distinguish
Hamiltonians acting on single  subsystems, called {\it local
Hamiltonians}, and Hamiltonians describing the coupling between two
or more systems, called {\it interaction (or entangling)
Hamiltonians}. In the unitary group, these Hamiltonians generate
local and entangling evolutions, respectively. In particular, if one
considers a multipartite system composed of $N$ subsystems of
dimensions $n_1, \ldots, n_N$, the space of all possible
Hamiltonians is given by $i{\mathfrak u}(n_1 n_2 \cdot \cdot \cdot
n_N)$. Once we have an orthogonal basis in $i{\mathfrak u}(n_j)$, $j
= 1, \ldots, N$, given by $H^j_{l_j}$, $l_j = 1, \ldots, n_j^2$,
then a basis of ${\mathfrak u}(n_1 n_2 \cdot \cdot \cdot  n_N)$ is
given by \be{hh} i H^1_{l_1} \otimes H^2_{l_2} \otimes \ldots
\otimes H^N_{l_N}. \ee The subalgebra of local Hamiltonians is
spanned by elements where all the factors in the tensor product are
equal to the identity except one. They produce unitary evolutions of
the factorized form $X_1 \otimes X_2 \otimes \ldots \otimes X_N$
which corresponds to transformations on the single subsystems. In
this spirit, the decomposition given in~\cite{KG} recursively
factorizes a unitary transformation on $n$ qubits in local and
entangling transformations, the latter one acting on two subsystems
at a time only. In a recent paper~\cite{Mikofigata} a method was
given to generate decompositions in the tensor product space for
general multipartite quantum systems of arbitrary dimensions
starting from decompositions of evolutions on the single subsystems.

In the spirit of the last two works cited, we present in this paper
a recursive procedure to decompose the unitary evolution of a
bipartite system {\it of arbitrary dimensions} so that every
evolution is factorized into simple terms and it is clear what the
entangling and local contributions of a single transformation are.
The procedure we present applies recursively the Cartan
decomposition of the Lie algebra ${\mathfrak{so}}(n)$ by keeping the
tensor product basis representation of the Lie algebra ${\mathfrak
u}(n)$. In this basis, at the end of the procedure, it is easy to
analyze the local and entangling character of each factor.

The paper is organized as follows: in the next section we review
some basic concepts and results on the Cartan classification of
symmetric spaces and Cartan decompositions. We shall mention only
the facts needed in the sequel of the paper. A detailed treatment
can be found in~\cite{Helgason}. The decomposition procedure we
describe consists of an initial step which reduces the problem to a
decomposition of the orthogonal group and a recursive procedure
which allows to decompose further the elements of the orthogonal
group. These steps are described in Sections~\ref{Init}
and~\ref{Recurs}, respectively. We give several remarks highlighting
the local and entangling factors needed in the factorization.
Section~\ref{Ultim} is devoted to a discussion and a numerical
example.

\vs

\noindent {\bf Notation:} In the following we will use several times
the definition of the elementary matrices $E_{mn}$,
\begin{equation} \label{AppA0}
(E_{mn})_{rs} = \delta_{mr} \delta_{ns},
\end{equation}
and of their antisymmetric and symmetric superposition, respectively
as
\begin{equation} \label{AppA0bis}
\Delta_{mn} := E_{mn} - E_{nm}, \quad \Omega_{mn} := E_{mn} +
E_{nm}.
\end{equation}
We denote by $A^T$ the transposed of the matrix $A$. A $n \times m$
rectangular matrix is denoted by $A_{n \times m}$; the $n \times n$
identity matrix by ${\bf 1}_n$. Finally, we call a {\it sign} matrix
a matrix of the form $diag(\pm 1, \pm 1, \ldots, \pm 1)$ with all
the possible combinations of $+$ and $-$.


\section{Background material}

In the following, we shall use (in a recursive manner) {\it Cartan
decompositions} of the Lie algebras ${\mathfrak{su}}(n)$ and
${\mathfrak{so}}(n)$ as well as Cartan decompositions of direct
products of (isomorphic copies of)  these Lie algebras. A Cartan
decomposition of a semisimple Lie algebra ${\mathfrak l}$ is a
vector space decomposition \be{RI1} {\mathfrak l} ={\mathfrak k}
\oplus {\mathfrak p}, \ee where ${\mathfrak k}$ is a subalgebra,
namely \be{RI2} [{\mathfrak k}, {\mathfrak k}] \subseteq {\mathfrak
k}, \ee and the complement subspace ${\mathfrak p} = {\mathfrak
k}^{\perp}$ is such that \be{RI3} [{\mathfrak p}, {\mathfrak p}]
\subseteq {\mathfrak k}, \qquad [{\mathfrak p}, {\mathfrak k}]
\subseteq {\mathfrak p}. \ee

To a Cartan decomposition of ${\mathfrak l}$ there corresponds a
factorization of $e^{\mathfrak l}$, the connected Lie group
associated to $\mathfrak l$, such that every element $X \in
e^{\mathfrak l}$ can be written as \be{RI5} X = K P,  \ee where $K$
belongs to $e^{\mathfrak k}$, the connected Lie group corresponding
to ${\mathfrak k}$, and $P$ is the exponential of an element in
${\mathfrak p}$. The coset space $e^{\mathfrak l}/e^{\mathfrak k}$
is called a {\it symmetric space} of $e^{\mathfrak l}$. A maximal
Abelian subalgebra of ${\mathfrak p}$ is called a {\it Cartan
subalgebra} associated to the decomposition and it is denoted by
${\mathfrak a}$. One can show, under appropriate assumptions, that
\be{RI6} \bigcup_{K \in e^{\mathfrak k}} K {\mathfrak a} K^{-1} =
{\mathfrak p},  \ee so that the factorization (\ref{RI5}) refines to
\be{RI10} X = K_1 A K_2, \ee with $K_1, K_2 \in e^{\mathfrak k}$ and
$A \in e^{\mathfrak a}$. The dimension of the Cartan subalgebra
${\mathfrak a}$ is called the {\it rank} of the decomposition (or of
the associated symmetric space).

\vs

Cartan has classified all the symmetric spaces of the classical Lie
groups, i.e. the Lie groups $SU(n)$, $Sp(n)$ and $SO(n)$, and has
shown that, up to conjugacy \footnote{A conjugacy on ${\mathfrak l}$
is a map $m : {\mathfrak l} \rightarrow {\mathfrak l}$ such that
$m(L) = M L M^{-1}$ for some $M \in e^{\mathfrak l}$}, the
corresponding decompositions fall in one of few classes which he has
described. In particular, for ${\mathfrak l}= {\mathfrak{su}}(n)$,
there are three types of decompositions labeled by AI, AII, AIII. In
the following,  we shall use only decompositions of the type AI,
given by \be{RI11} {\mathfrak k} = {\mathfrak{so}}(n), \qquad
{\mathfrak p} = {\mathfrak{so}}(n)^{\perp}. \ee The inner product in
the space ${\mathfrak{su}}(n)$ is given by $\langle A, B \rangle =
Tr(AB^*)$ so that ${\mathfrak{so}}(n)^{\perp}$ is the subspace of
${\mathfrak{su}}(n)$ spanned by purely imaginary matrices. The rank
of this decomposition is $n - 1$. Decompositions of the type AII and
AIII will not be considered here.

We consider now the Lie algebra ${\mathfrak l} =
{\mathfrak{so}}(n)$. If $n = 1$, this algebra contains only the null
matrix ${\bf 0}$. For $n > 2$, when $n$ is odd there is only one
type of Cartan decomposition, denoted by BDI. Fixing two positive
integers $r$ and $q$ such that $r \geqslant q \geqslant 1$ and $r +
q = n$, the matrices $k \in {\mathfrak k}$ have the form \be{RI14} k
= \begin{pmatrix}
                             A & 0 \\
                             0 & B
\end{pmatrix}, \ee with $A \in {\mathfrak{so}}(r)$ and $B \in {\mathfrak{so}}(q)$.
Matrices $p \in {\mathfrak p}$ have the form \be{RI15} p =
\begin{pmatrix}
                             0 & C \\
                             -C^T & 0
\end{pmatrix}, \ee for a general $r \times q $ matrix $C$.
The rank of this decomposition is $q$. In this paper, we will not
consider other decompositions.


\section{Decomposition of unitary evolutions  in $U(d_1d_2)$; Initial step}
\label{Init}

Consider two interacting quantum systems ${\cal S}_1$ and ${\cal
S}_2$ whose associated Hilbert spaces have dimensions $d_1$ and
$d_2$, respectively. According to the procedure described
in~\cite{Mikofigata}, it is possible to obtain a decomposition for
${\mathfrak{su}}(d_1 d_2)$ from decompositions of type AI of the Lie
algebras associated to each subsystem, i.e. ${\mathfrak{su}}(d_1)$
and ${\mathfrak{su}}(d_2)$. We also include in the algebra scalar
matrices, physically corresponding to shifts in the energy. We write
\be{R1} {\mathfrak u}(d_1) = {\mathfrak{so}}(d_1) \oplus
{\mathfrak{so}}(d_1)^\perp,  \qquad  {\mathfrak u}(d_2) =
{\mathfrak{so}}(d_2) \oplus {\mathfrak {so}}(d_2)^\perp. \ee Let
$\sigma^{j}$, $j = 1, 2$, be a generic element of  an orthogonal
basis of $i{\mathfrak{so}}(d_j)$, and $S^{j}$, $j = 1, 2$,  a
generic element of an orthogonal basis of
$i{\mathfrak{so}}(d_j)^\perp$. Then the subalgebra of ${\mathfrak
u}(d_1d_2)$ defined by \be{R3} {\mathfrak k} := span \{i \sigma^1
\otimes S^2, iS^1 \otimes \sigma^2 \}, \ee along with its orthogonal
complement in ${\mathfrak u}(d_1d_2)$, \be{R4} {\mathfrak p} := span
\{ i \sigma ^1 \otimes \sigma^2, i S^1 \otimes S^2\}, \ee define a
Cartan decomposition of ${\mathfrak u}(d_1d_2)$ as \be{R5}
{\mathfrak u}(d_1d_2)= {\mathfrak k} \oplus {\mathfrak p}. \ee This
decomposition is of type AI~\cite{Mikofigata} as ${\mathfrak k}$ is
conjugate to ${\mathfrak{so}}(d_1d_2)$ and ${\mathfrak p}$ to
${\mathfrak{so}}(d_1d_2)^\perp$. The rank of this decomposition is
$d_1d_2$. A basis of the maximal abelian subalgebra ${\mathfrak a}
\subseteq {\mathfrak p}$ is given by tensor products of elements of
the orthogonal basis of the maximal Abelian subalgebras associated
to the single subsystems, which are of dimensions $d_1$ and $d_2$
respectively. Denoting by $D^1$ the diagonal elements of the type
$S^1$, and by $D^2$ those of the type $S^2$, $\mathfrak a$ is given
by \be{R6} {\mathfrak a} := span \{ iD^1 \otimes D^2 \}. \ee The
associated Cartan factorization of $X \in U(d_1 d_2)$ is \be{R7} X =
K_1 A K_2 \ee according to the notation of the previous section.

\vs

\br{localita1} Only in the simplest case of the decomposition of
${\mathfrak{su}}(4)$ (i.e. $d_1=2$ and $d_2=2)$, studied for example
in \cite{Zhang}, the decomposition (\ref{R7}) is  a decomposition in
local and nonlocal transformations. The local transformations are
products of exponentials of matrices of the form $iH \otimes {\bf
1}$ or ${\bf 1} \otimes iH$, where ${\bf 1}$ is the identity matrix
of appropriate dimensions and $H$ is a generic matrix in
$i{\mathfrak u}(d_1)$ or $i{\mathfrak u}(d_2)$. Both local and
nonlocal transformations are possibly present in the $K_1$ and $K_2$
factors as well as in the $A$ factor. However, obtaining a
decomposition in terms of tensor product matrices will allow us to
identify exactly where the local and nonlocal transformations are
present in the final transformation. \er

\br{localita2} We notice that, in general, only one nonlocal
transformation, along with the set of the local transformations, is
sufficient to obtain all the possible values for $A$ in (\ref{R7}).
To see this, notice that the factor $A$ is the finite product of
exponentials of elements of the form $i \alpha_{jk} E_{jj} \otimes
E_{kk}$, with $j = 1, \ldots, d_1$, $k = 1, \ldots, d_2$, and
$\alpha_{jk}$ real numbers. Since, for every $l$,  $E_{ll}$ is
unitarily equivalent to $E_{11}$, the Hamiltonian $H =  E_{11}
\otimes E_{11}$, along with local transformations, is sufficient to
generate any element of the form $A$. Notice that an alternative
(universal) {\it nonlocal Hamiltonian} is given by an Ising
interaction between two spins, which in our notation reads as
$(E_{11} - E_{d_1 d_1}) \otimes (E_{11} - E_{d_2 d_2})$. \er

We now turn our attention to decomposing the elements $K_1$ and
$K_2$ in (\ref{R7}). This will be obtained through a recursive
procedure via iterate decompositions of ${\mathfrak {so}}(n)$.


\section{Decomposition of unitary evolutions  in
$U(d_1d_2)$; Recursive procedure} \label{Recurs}

The Lie algebra ${\mathfrak k}$ defined in (\ref{R3}) is conjugate
to ${\mathfrak{so}}(d_1d_2)$. We rewrite its definition below:
\begin{equation*}
{\mathfrak k} := span \{i \sigma^1 \otimes S^2,
iS^1 \otimes \sigma^2 \},
\end{equation*}
with $\sigma^j$, $j = 1, 2$, belonging to an orthogonal basis of
$i {\mathfrak{so}}(d_j)$ and $S^j$, $j = 1, 2$, belonging to an
orthogonal basis of  $i {\mathfrak{so}}(d_j)^\perp $.

A special case arises when $d_1 = d_2 = 1$, and only the matrix $\bf
0$ belongs to the corresponding Lie algebra. This case is not of
physical interest  as it would imply a one dimensional quantum
system. However it may arise as the final step of the recursive
procedure we are going to present. A special, nonphysical case is
also $d_1 = 2$ and $d_2 = 1$ or viceversa.  In this case, the Lie
algebra ${\mathfrak k}$ contains only one element. Another special
case is given by $d_1 = d_2 = 2$. In this case, consider the Pauli
matrices \be{R9} \sigma_x := \begin{pmatrix}
                             0 & 1 \\
                             1 & 0
\end{pmatrix}, \qquad \sigma_y := \begin{pmatrix}
                             0 & -i \\
                             i & 0
\end{pmatrix}, \qquad \sigma_z := \begin{pmatrix}
                             1 & 0 \\
                             0 & -1
\end{pmatrix}, \ee and the $2 \times 2$ identity matrix ${\bf 1}$.
Then ${\mathfrak k} = {\mathfrak{so}}(4)$ is the direct sum of two
commuting subalgebras, ${\mathfrak s}_1$ and ${\mathfrak s}_2$, each
isomorphic to ${\mathfrak{so}}(3)$, and given by
\begin{eqnarray}\label{R10}
{\mathfrak s}_1 &:=& span \{ i \sigma_y \otimes {\bf 1}, i \sigma_x
\otimes \sigma_y, i \sigma_z \otimes \sigma_y \}, \nonumber \\
{\mathfrak s}_2 &:=& span \{ i {\bf 1} \otimes \sigma_y, i \sigma_y
\otimes \sigma_x, i \sigma_y \otimes \sigma_z \}.
\end{eqnarray}
Therefore $K_1$ (and analogously $K_2$) in (\ref{R7}) can be written
as the product \be{R12} K_1 = F_1 F_2 = F_2 F_1, \ee with $F_1$ and
$F_2$ in the Lie group corresponding to ${\mathfrak s}_1$ and
${\mathfrak s}_2$ respectively. A Cartan decomposition can be
performed on ${\mathfrak s}_1$ (and analogously on ${\mathfrak
s}_2$) which is an Euler decomposition as ${\mathfrak s}_1$ is
isomorphic to ${\mathfrak{so}}(3)$, and allows to express $F_1$ as
\be{R13} F_1 = L_1 N L_2,  \ee with $L_j = e^{\alpha_j i \sigma_y
\otimes {\bf 1}}$, $j = 1, 2$, for real parameters $\alpha_j$, and
$N = e^{\beta i \sigma_x \otimes \sigma_y}$ for a real parameter
$\beta$. Notice that $L_1$ and $L_2$ are local transformations while
$N$ is nonlocal. The same can be done for $F_2$, moreover the
nonlocal transformation for $F_2$ can be obtained using a local
similarity transformation from the one for $F_1$ or viceversa, so
that for $F_1$ and $F_2$ we need only one nonlocal Hamiltonian.

\vs

Consider now the case where at least one between $d_1$ and $d_2$ is
greater than $2$. As it was done for the initial step in the
previous section, we look for decompositions concerning the single
subsystems to induce a decomposition on the total bipartite system.
The elements of the type $i \sigma^1$ and $i \sigma^2$ are real,
skew-symmetric, square matrices of dimensions $d_1$ and $d_2$
respectively. Let $d_1 > 2$ without loss of generality. On the Lie
algebra of $d_1 \times d_1$ skew-symmetric matrices (that is,
matrices of the type $\sigma^1$) we perform a decomposition of the
type BDI (see previous section) by selecting two positive integers
$r_1 \geqslant q_1 \geqslant 1$ so that $r_1 + q_1 = d_1$. In the
resulting Cartan decomposition, ${\mathfrak{so}}(d_1) = {\mathfrak
k} \oplus {\mathfrak p}$, the Lie algebra ${\mathfrak k}$ is spanned
by block diagonal skew-symmetric matrices with the upper block of
dimension $r_1$ and the lower block of dimension $q_1$. We denote
this type of matrices by $i \sigma^{1,D}$ (where $D$ stands for
`diagonal'). The skew-symmetric matrices in the complement
${\mathfrak p}$ will be denoted by $i \sigma^{1,A}$ (where $A$
stands for `anti-diagonal'). We as well separate matrices of the
type $S^1$ into block diagonal and block anti-diagonal and denote
them by $iS^{1,D}$ and $iS^{1,A}$, respectively. Analogously, we
define a decomposition of type BDI on ${\mathfrak{so}}(d_2)$
introducing two positive integers $r_2 \geqslant q_2 \geqslant 1$,
with $r_2 + q_2 = d_2$ and matrices of the type $i \sigma^{2,D}$,
$i\sigma^{2,A}$, $iS^{2,D}$ and $iS^{2,A}$. In the special case
where $d_2 = 2$, we can only choose $r_2 = q_2 = 1$ and we do not,
in fact, obtain a decomposition of ${\mathfrak{so}}(d_2)$ of the
type BDI. However, we still formally decompose matrices of the form
$\sigma^2$ and $S^2$ in (block) diagonal and (block) anti-diagonal
components and notice that, in this case,  the only matrix of the
type $i\sigma^{2,D}$ and $iS^{2,A}$ is the $2 \times 2$ zero matrix.

\vs

These decompositions on the two subsystems induce a decomposition on
the overall bipartite system. More precisely we decompose
${\mathfrak k}$ in (\ref{R3}) as follows \be{R15} {\mathfrak
k}:={\mathfrak k}^{\prime} \oplus {\mathfrak p}^{\prime}, \ee with
\begin{eqnarray}\label{R16}
{\mathfrak k}^{\prime} = span \{i \sigma^{1,D} \otimes S^{2,D}, i
S^{1,D} \otimes \sigma^{2,D}, i \sigma^{1,A} \otimes S^{2,A}, i
S^{1,A} \otimes \sigma^{2,A} \}, \nonumber \\ {\mathfrak p}^{\prime}
= span \{ i \sigma^{1,A} \otimes S^{2,D}, i \sigma^{1,D} \otimes
S^{2,A}, i S^{1,D}\otimes \sigma^{2,A}, i S^{1,A} \otimes
\sigma^{2,D} \}.
\end{eqnarray}

\vs

The following Theorem summarizes the features of this decomposition.
It also gives, in its proof, a coordinate transformation to write
the elements of the subalgebra ${\mathfrak k}^{\prime}$ and its
complement ${\mathfrak p}^{\prime}$ in the standard form.

\vs

\bt{Feat} The decomposition of ${\mathfrak k}$ defined in
(\ref{R15}),(\ref{R16}) is a Cartan decomposition, i.e.
\begin{equation*}
[{\mathfrak k}^{\prime}, {\mathfrak k}^{\prime}] \subseteq
{\mathfrak k}^{\prime}, \quad [{\mathfrak k}^{\prime}, {\mathfrak
p}^{\prime}] \subseteq {\mathfrak p}^{\prime}, \quad [{\mathfrak
p}^{\prime}, {\mathfrak p}^{\prime}] \subseteq {\mathfrak
k}^{\prime}.
\end{equation*}
As a decomposition of ${\mathfrak so}(d_1 d_2)$, it is a Cartan
decomposition of type BDI with indices $r$ and $q$ satisfying $r
\geqslant q \geqslant 1$, $r + q = d_1 d_2$, and with \be{R18bis} r
= r_1 r_2 + q_1 q_2, \qquad q = r_1 q_2 + q_1 r_2. \ee Accordingly,
the dimension of the associated Cartan subalgebra ${\mathfrak
a}^{\prime} \subseteq {\mathfrak p}^{\prime}$ is $q = r_1 q_2 + q_1
r_2$.

\et

\bpr We explicitly exhibit a  conjugacy which transforms elements of
${\mathfrak k}^{\prime}$ into the form (\ref{RI14}) and elements of
${\mathfrak p}^{\prime}$ into the form (\ref{RI15}). In particular,
notice that the matrices $i\sigma^{1,D} \otimes S^{2,D}$, $iS^{1,D}
\otimes \sigma^{2,D}$ have the form \be{forma1} k_1 :=
\begin{pmatrix}
                             A_{r_1 \times r_1} \otimes C_{r_2 \times r_2} & 0& 0 & 0 \\
                             0 & A_{r_1 \times r_1} \otimes D_{q_2\times q_2} & 0 & 0 \\
                             0 & 0 & B_{q_1 \times q_1} \otimes C_{r_2 \times r_2} & 0 \\
                             0 & 0 & 0 & B_{q_1 \times q_1} \otimes D_{q_2 \times q_2}
\end{pmatrix},
\ee while the matrices $i \sigma^{1,A} \otimes S^{2,A}$ and
$iS^{1,A} \otimes \sigma^{2,A}$ are of the form \be{forma2} k_2:=
\begin{pmatrix}
                             0 & 0 & 0 & F_{r_1 \times q_1} \otimes G_{r_2 \times q_2} \\
                             0 & 0 & \pm F_{r_1 \times q_1} \otimes G^T_{q_2 \times r_2}& 0  \\
                             0 & \mp F^T_{r_1 \times q_1} \otimes G_{r_2 \times q_2} & 0 & 0\\
                             -F^T_{q_1 \times r_1} \otimes G^T_{q_2 \times r_2} & 0 & 0 & 0
\end{pmatrix}.
\ee A straightforward calculation shows that, defining \be{conjugac}
R := \begin{pmatrix}
                             {\bf 1}_{r_1 r_2} & 0 & 0 & 0\\
                             0 & 0 & 0 & {\bf 1}_{q_1 q_2}\\
                             0 & {\bf 1}_{r_1 q_2} & 0 & 0 \\
                             0 & 0 & {\bf 1}_{r_2 q_1} & 0
\end{pmatrix},
\ee the matrices \be{transf1} \tilde k_1:= R k_1 R^T, \qquad \tilde
k_2 := R k_2 R^T, \ee have the form given in (\ref{RI14}) where the
upper block has dimension $r = r_1 r_2 + q_1 q_2$ and the lower
block has dimension $q = r_1 q_2 + q_1 r_2$. Analogously one shows
that the conjugacy defined in (\ref{conjugac}) transforms elements
in ${\mathfrak p}^{\prime}$ into elements of the form $p$ in
(\ref{RI15}).

\epr

\noindent In view of the decomposition (\ref{R15}) any element
$K_1$ (and analogously for $K_2$) in (\ref{R7}) can be written as
\be{R27} K_1=K_1^{\prime} A^{\prime} K_2^{\prime}, \ee where
$K_1^{\prime}$ and $K_2^{\prime}$ belong to the Lie group
associated to the Lie algebra ${\mathfrak k}^{\prime}$, conjugate
to ${\mathfrak {so}}(r) \oplus {\mathfrak {so}}(q)$, and are to be
further factorized. The matrix $A^{\prime}$ belongs to the Abelian
Lie subgroup associated to the maximal Abelian subalgebra
${\mathfrak a}^{\prime} \in {\mathfrak p}^{\prime}$. In the
following Proposition we find an orthogonal basis for such a
Cartan subalgebra expressing it in terms of tensor
products\footnote{ An alternative procedure is to transform the
Lie algebra ${\mathfrak k}'$ according   to the change of
coordinates (\ref{conjugac}) and
 to find the Cartan subalgebra in the standard basis}.

\bp{Figata1}

The $(r_1 + q_1) q_2$ matrices \be{R30} N_{jk} := E_{jj} \otimes
\Delta_{k, r_2 + k}, \qquad j = 1, \ldots, r_1 + q_1, \quad k = 1,
\ldots, q_2, \ee along with the $(r_2 - q_2) q_1$ matrices \be{R31}
M_{fl} := \Delta_{f, r_1 + f} \otimes E_{ll}, \qquad l = q_2 + 1,
\ldots, r_2, \quad f = 1, \ldots, q_1, \ee span a Cartan subalgebra
${\mathfrak a}^{\prime} \in {\mathfrak p}^{\prime}$. \ep

\bpr The dimension of the vector space spanned by the matrices in
(\ref{R30}), (\ref{R31}) is, in fact, $(r_1 + q_1) q_2 + (r_2 - q_2)
q_1 = r_1 q_2 + q_1 r_2$. Therefore, we only have to verify that
matrices of the type (\ref{R30}) and (\ref{R31}) commute with each
other. The commutator between two matrices of the type (\ref{R30})
always vanishes. Analogously, matrices of the type (\ref{R31})
commute with each other. The Lie bracket of matrices of the type
(\ref{R30}) and (\ref{R31}) vanishes too, since the products of
matrices $\Delta_{k, r_2 + k}$ and $E_{ll}$ in the second factors
are always zero. \epr

\vs

\br{Ancent}

It follows from Proposition \ref{Figata1} that the element
$A^{\prime}$ in (\ref{R27}) is the exponential of a linear
combination of matrices (\ref{R30}) and (\ref{R31}) or
(equivalently) the product of exponentials of matrices proportional
to these. The resulting unitary transformations may be entangling or
local. However, since all the matrices  of the form $E_{jj}$ are
unitarily equivalent to each other and the matrices of the type
$\Delta_{kl}$ are also unitarily equivalent to each other, only one
(entangling) Hamiltonian of the type (\ref{R30}), one of the type
(\ref{R31}) along with local operations are sufficient (and
necessary) to generate all the possible factors $A^{\prime}$ in
(\ref{R27}).

\er

\br{Furtherred} A further reduction of the nonlocal Hamiltonians to
be used is obtained by noticing that all the transformations in
Remark \ref{localita2} and in Proposition \ref{Figata1} can be
obtained with only one  Ising Hamiltonian and local transformations.
Therefore only one nonlocal Hamiltonian  is needed to implement all
of these transformations.\er

\vs

We now further factorize the elements $K_1^{\prime}$ and
$K_2^{\prime}$ in (\ref{R27}) using, once again, a Cartan
decomposition of the Lie algebra ${\mathfrak k}^{\prime}$ isomorphic
to ${\mathfrak {so}}(r) \oplus {\mathfrak {so}}(q)$, with $r$ and
$q$ defined in~(\ref{R18bis}). In particular, we decompose
${\mathfrak k}^{\prime}$ as follows

\be{R39} {\mathfrak k}^{\prime}= {\mathfrak k}^{\prime \prime}
\oplus {\mathfrak p}^{\prime \prime}, \ee with
\begin{eqnarray}\label{R40} {\mathfrak k}^{\prime \prime} = span \{ i
\sigma^{1,D} \oplus S^{2,D}, S^{1,D} \oplus i \sigma^{2,D} \},
\nonumber \\
{\mathfrak p}^{\prime \prime}= span \{ i  \sigma^{1,A} \otimes
S^{2,A}, S^{1,A} \otimes i \sigma^{2,A} \}. \end{eqnarray} The
matrices in ${\mathfrak k}^{\prime \prime}$ are block diagonal
matrices and ${\mathfrak k}^{\prime \prime}$  is \be{R42} {\mathfrak
k}^{\prime \prime} = {\mathfrak {so}}(r_1 r_2) \oplus {\mathfrak
{so}}(r_1 q_2) \oplus {\mathfrak {so}}(q_1 r_2) \oplus {\mathfrak
{so}}(q_1 q_2), \ee where each term refers to a block on the
diagonal. For example, the first block, corresponding to ${\mathfrak
{so}}(r_1r_2)$, contains matrices obtained as tensor products $i
\sigma^{1,D} \otimes S^{2,D}$ or $S^{1,D} \otimes \sigma^{2,D}$
where all matrices involved have the second block equal to zero.
This corresponds to two decompositions of the type BDI, one on
${\mathfrak {so}}(r)$ and the other on ${\mathfrak {so}}(q)$. The
Cartan subalgebra in ${\mathfrak p}^{\prime \prime}$ is the direct
sum of the two Cartan subalgebras of the two decompositions. It has
dimension $q_1 q_2 + min \{r_1 q_2, q_1 r_2 \}$. The following
proposition explains how to find a basis of this Cartan subalgebra
as tensor product matrices\footnote{Alternatively, one can construct
a basis for this Cartan subalgebra working in the standard
representation using the coniugacy given in (\ref{conjugac})}.

\bp{PropoCar2} A Cartan subalgebra of the decomposition (\ref{R39}),
(\ref{R40}) is spanned  by the $q_1 q_2$ matrices \be{R44} N_{jm,ln}
:= \begin{pmatrix}
                             0 & E_{jl} \\
                             -E_{jl}^T & 0
\end{pmatrix} \otimes \begin{pmatrix}
                             0 & E_{mn} \\
                             E_{mn}^T & 0
\end{pmatrix} + \begin{pmatrix}
                             0 & E_{jl} \\
                             E_{jl}^T & 0
\end{pmatrix} \otimes \begin{pmatrix}
                             0 & E_{mn} \\
                             -E_{mn}^T & 0,
\end{pmatrix} \ee with  $1 \leqslant j \leqslant r_1$, $1\leqslant l \leqslant
q_1$ and $1 \leqslant m\leqslant r_2$, $1 \leqslant n \leqslant q_2$
satisfying \be{R45} (j - 1) r_2 + m = s, \quad (l - 1) q_2 + n = s
\ee with $s = 1, \ldots, q_1 q_2$, along with matrices \be{R46}
M_{jm,ln} := \begin{pmatrix}
                             0 & E_{jl} \\
                             -E_{jl}^T & 0
\end{pmatrix} \otimes \begin{pmatrix}
                             0 & E_{mn} \\
                             E_{mn}^T & 0
\end{pmatrix} - \begin{pmatrix}
                             0 & E_{jl} \\
                             E_{jl}^T & 0
\end{pmatrix} \otimes \begin{pmatrix}
                             0 & E_{mn} \\
                             -E_{mn}^T & 0
\end{pmatrix}, \ee with  $1 \leqslant j \leqslant r_1$, $1 \leqslant l \leqslant
q_1$ and $1 \leqslant m \leqslant r_2$, $1 \leqslant n \leqslant
q_2$ satisfying \be{R47} (j - 1) q_2 + n = s, \quad (l - 1) r_2 + m
= s \ee with $s = 1, \ldots, min \{ r_1 q_2, q_1 r_2 \}$. \ep
\br{Remindici} For all $s$, there corresponds a unique pair $(l, n)$
so that the second relation in (\ref{R45}) is verified. There is
some freedom in choosing the pairs $(j, m)$ satisfying the first
relation in (\ref{R45}). However, for  every value of $s$, and
therefore of $l$ and $n$, one is allowed to choose a {\it unique}
pair $(j, m)$. An analogous meaning has the notation in (\ref{R46})
and (\ref{R47}). \er

\bpr Matrices of the form (\ref{R44}) commute with matrices of the
form (\ref{R46}), since these matrices form Cartan subalgebras
associated to decompositions of ${\mathfrak {so}}(r)$ and
${\mathfrak {so}}(q)$, respectively. To show that matrices of type
(\ref{R44}) commute, one verifies that the commutators of two
matrices corresponding to indices $(j_1 m_1, l_1 n_1)$ and $(j_2
m_2, l_2 n_2)$ vanish. In fact all the blocks of such matrices are
zero except for the $1,1$ and $2,2$ blocks which, from a direct
calculation, turn out to be equal to \be{R50} 4(E_{j_2j_1}
\delta_{l_1 l_2} \otimes E_{m_2 m_1} \delta_{n_1 n_2}- E_{j_1 j_2}
\delta_{l_1 l_2} \otimes E_{m_1 m_2} \delta_{n_1 n_2}), \ee and
\be{R51} 4(E_{l_2 l_1} \delta_{j_1 j_2} \otimes E_{n_2 n_1}
\delta_{m_1 m_2} - E_{l_1l_2} \delta_{j_1 j_2} \otimes E_{n_1 n_2}
\delta_{m_1 m_2}), \ee respectively.  However these are also zero if
$n_1 \ne n_2$ and/or $l_1 \ne l_2$ as well as  in the case $l_1 =
l_2$, $n_1 = n_2$ (and therefore $j_1 = j_2$, $m_1 = m_2$, see
Remark \ref{Remindici}). A perfectly analogous argument holds in the
case of commutators of matrices of the form (\ref{R46}). \epr

\br{localita5} Notice that all the Hamiltonians (\ref{R44}) are
locally unitarily equivalent to each other. The same is true for the
Hamiltonians (\ref{R46}). Therefore only two more entangling
Hamiltonians are needed.

\er

At this point we are left with the Lie algebra ${\mathfrak
{so}}(p_1p_1)\oplus {\mathfrak {so}}(q_1 q_2) \oplus {\mathfrak
{so}}(p_1 q_2) \oplus {\mathfrak {so}}(p_2 q_1)$. The construction
proceeds recursively by decomposing each one of the four component
Lie algebras and so on, until one finds one of the Lie algebras
${\mathfrak {so}}(1)$ (which we define as the element zero),
${\mathfrak {so}}(2)$ (which consists of a single element),
${\mathfrak {so}}(3)$ or ${\mathfrak {so}}(4)$ (which are  treated
as it was explained at the beginning of the procedure).


\section{Discussion and a numerical example}
\label{Ultim}

In order to illustrate the Lie group decomposition described in the
previous sections, we consider the {\it generalized SWAP} operator
$X_{sw}$ acting on three qubits and rotating their states in a
cyclic fashion. Its action is defined in the tensor product basis as
\begin{equation}\label{swap}
    X_{sw}: \vert i \rangle_1 \otimes \vert j \rangle_2 \otimes \vert k
    \rangle_3 \rightarrow \vert k \rangle_1 \otimes \vert i \rangle_2
    \otimes \vert j \rangle_3,
\end{equation}
where $i, j, k = 0, 1$ and $\{ \vert 0 \rangle, \vert 1 \rangle
\}_{1,2,3}$ are orthonormal basis for the Hilbert spaces of the
three systems.

This operator is relevant in quantum information and computation
since it enables to switch the quantum states of different systems.
For example, assume that one is interested in the state of the third
system, but only the first system is accessible and can be
controlled; then the  application of the generalized SWAP operator
will enable to transfer the state of the third system to the first
system. As local operations alone clearly cannot implement the
generalized SWAP and this has to involve some degree of entanglement
among the various subsystems. We consider an hypothetical situation
where it is possible to create interaction between the first qubit
and the other two as a whole although it is difficult to create
interactions with the single qubits $2$ and $3$. This justifies to
consider the total Hilbert space as the tensor product of a
2-dimensional subspace with a 4-dimensional one (that is, $d_1 = 2$
and $d_2 = 4$). Therefore $X_{sw} \in U (8)$ will be decomposed
accordingly \footnote{We believe that extensions of the procedure
presented here to multipartite systems are possible at the price of
an increased notational complexity.}.

In the specified basis, with standard ordering, the matrix
representation of this operator is given by
\begin{equation}\label{xsw}
    X_{sw} = \begin{pmatrix}
               1 & 0 & 0 & 0 & 0 & 0 & 0 & 0 \\
               0 & 0 & 1 & 0 & 0 & 0 & 0 & 0 \\
               0 & 0 & 0 & 0 & 1 & 0 & 0 & 0 \\
               0 & 0 & 0 & 0 & 0 & 0 & 1 & 0 \\
               0 & 1 & 0 & 0 & 0 & 0 & 0 & 0 \\
               0 & 0 & 0 & 1 & 0 & 0 & 0 & 0 \\
               0 & 0 & 0 & 0 & 0 & 1 & 0 & 0 \\
               0 & 0 & 0 & 0 & 0 & 0 & 0 & 1 \\
             \end{pmatrix}.
\end{equation}
This transformation belongs to $SO (8)$, therefore the first step of
the decomposition of $U (8)$ is trivial: $K_1 = X_{sw}$ and $A = K_2
= {\bf 1}$.

For the first step of the recursive part of the procedure, we choose
$r_1 = q_1 = 1$ and $r_2 = q_2 = 2$. We find it convenient to work
in the basis of the Hilbert space such that ${\mathfrak k}' =
{\mathfrak {so}} (2) \oplus {\mathfrak {so}} (4)$, obtained by
performing the change of basis $R \vert ijk \rangle \rightarrow
\vert ijk \rangle^{\prime}$, with $R$ given in (\ref{conjugac}),
which,  in this particular case ($r_1 = q_1 = 1$, $r_2 = q_2 = 2$),
takes the form
\begin{equation}\label{cob}
    R = \begin{pmatrix}
               {\bf 1}_{2} & 0 & 0 & 0\\
               0 & 0 & 0 & {\bf 1}_{2} \\
               0& {\bf 1}_{2} & 0 & 0 \\
               0&0&{\bf 1}_{2}& 0 \\
        \end{pmatrix}.
\end{equation}
In these coordinates, the SWAP operator is written  as
$\tilde{X}_{sw} = R X_{sw} R^{T}$, that is
\begin{equation}\label{coba}
    \tilde{X}_{sw} = \begin{pmatrix}
               1 & 0 & 0 & 0 & 0 & 0 & 0 & 0 \\
               0 & 0 & 0 & 0 & 1 & 0 & 0 & 0 \\
               0 & 0 & 0 & 0 & 0 & 0 & 0 & 1 \\
               0 & 0 & 0 & 1 & 0 & 0 & 0 & 0 \\
               0 & 0 & 0 & 0 & 0 & 0 & 1 & 0 \\
               0 & 0 & 1 & 0 & 0 & 0 & 0 & 0 \\
               0 & 1 & 0 & 0 & 0 & 0 & 0 & 0 \\
               0 & 0 & 0 & 0 & 0 & 1 & 0 & 0 \\
             \end{pmatrix}.
\end{equation}
The elements of the Cartan subalgebra $\mathfrak {a}^{\prime}$
(defined in Proposition~\ref{Figata1}), which in this case are only
of the form (\ref{R30}), are transformed
by $R$  into elements of the form $$ \tilde{a}^{\prime} = \begin{pmatrix} 0 & D \\
-D & 0 \end{pmatrix}.$$ The computational problem is to find two $4
\times 4$ diagonal matrices, $D_1$ and $D_2$, with $$\tilde{A}^{\prime} = e^{\tilde{a}^{\prime}} = \begin{pmatrix}D_1 & D_2 \\
-D_2& D_1\\\end{pmatrix},$$ $\tilde{A}^{\prime} \in SO(8)$, and
matrices $K_{ij} \in SO (4)$, $i, j = 1, 2$, such that
\begin{equation}\label{secxsw}
    \tilde{K}_1^{\prime} = \begin{pmatrix}
                       K_{11} & 0 \\
                       0 & K_{12} \\
                     \end{pmatrix},
                     \qquad
     \tilde{K}_2^{\prime} = \begin{pmatrix}
                       K_{21} & 0 \\
                       0 & K_{22} \\
                     \end{pmatrix}
\end{equation}
and $\tilde{X}_{sw} = \tilde{K}_1^{\prime} \tilde{A}^{\prime}
\tilde{K}_2^{\prime}$. To perform this task we propose an  algorithm
which uses ideas similar to the ones for other Cartan decompositions
(cf. e.g. \cite{BBO1} and the references therein).  We illustrate
this algorithm for the dimensions of our problem but generalizations
to other dimensions are obvious. Let us write $\tilde X_{sw}$ with
$4 \times 4$ blocks $\tilde X_{ij}$, $i,j = 1,2$, as \be{blocchi}
\tilde X_{sw}=\begin{pmatrix}\tilde X_{11} & \tilde X_{12} \cr
\tilde X_{21} & \tilde X_{22}. \end{pmatrix}.   \ee Equation
(\ref{secxsw}) is equivalent to the four matrix equations
\begin{eqnarray}
\tilde X_{11} &=& K_{11}D_1K_{21}, \label{prima}\\
\tilde X_{12} &=& K_{11} D_2 K_{22}, \label{seconda} \\
\tilde X_{21} &=& -K_{12} D_2 K_{21}, \label{terza} \\
\tilde X_{22} &=& K_{12} D_1 K_{22}. \label{quarta}
\end{eqnarray}
From the first one, we obtain \be{eigeq} \tilde X_{11} \tilde
X_{11}^T K_{11}=K_{11} D_1^2,  \ee which is an eigenvalue equation
as $D_1^2$ is diagonal. In the generic case, when all the
eigenvalues of $\tilde X_{11} \tilde X_{11}^T$ are different,
equation (\ref{eigeq}) determines $K_{11}$ and $D_1$ up to the right
product by a sign matrix (and the fact that $det(K_{11}) = 1$).
Moreover, $D_1$ gives  $D_2$ up to a sign matrix from the
requirement that $D_1^2 + D_2^2 = {\bf 1}_{4}$. Using $K_{11}$ and
$D_2$ in (\ref{seconda}), we obtain $K_{22}$ up to a sign matrix.
Plugging  $K_{22}$ in (\ref{quarta}), we get $K_{12}$ up to a sign
matrix and from (\ref{terza}) we find $K_{21}$. Finally, we adjust
the sign matrices to make (\ref{prima}) through (\ref{quarta})
consistently verified. In the case where $\tilde X_{11} \tilde
X_{11}^T$ has multiple eigenvalues, there is more freedom in the
choice of $K_{11}$ at the initial step, but then one proceeds in the
same way and determines the other matrices up to some degree of
freedom. At the end of the procedure, these degrees of freedom are
exploited  to make (\ref{prima})-(\ref{quarta}) jointly satisfied.

\vs

\noindent Using this procedure, we have found for our example
\begin{eqnarray}\label{kij}
    K_{11} &= \begin{pmatrix}
               1 & 0 & 0 & 0 \\
               0 & 0 & 1 & 0 \\
               0 & 1 & 0 & 0 \\
               0 & 0 & 0 & -1 \\
             \end{pmatrix}, \qquad
             K_{12} &= - \begin{pmatrix}
               0 & 0 & 0 & 1 \\
               0 & 0 & 1 & 0 \\
               0 & 1 & 0 & 0 \\
               1 & 0 & 0 & 0 \\
             \end{pmatrix}, \nonumber \\
             K_{21} &= \begin{pmatrix}
               1 & 0 & 0 & 0 \\
               0 & 1 & 0 & 0 \\
               0 & 0 & 1 & 0 \\
               0 & 0 & 0 & 1 \\
             \end{pmatrix}, \qquad
             K_{22} &= \begin{pmatrix}
               0 & -1 & 0 & 0 \\
               0 & 0 & 0 & 1 \\
               1 & 0 & 0 & 0 \\
               0 & 0 & 1 & 0 \\
             \end{pmatrix}, \\
             D_1 &= \begin{pmatrix}
               1 & 0 & 0 & 0 \\
               0 & 0 & 0 & 0 \\
               0 & 0 & 0 & 0 \\
               0 & 0 & 0 & -1 \\
             \end{pmatrix}, \qquad
             D_2 &= \begin{pmatrix}
               0 & 0 & 0 & 0 \\
               0 & 1 & 0 & 0 \\
               0 & 0 & 1 & 0 \\
               0 & 0 & 0 & 0 \\
             \end{pmatrix}. \nonumber
\end{eqnarray}
We can repeat the same procedure as above to further factorize
$K_{11}$, $K_{12}$, $K_{21}$, $K_{22}$, and to obtain
$\tilde{K}_1^{\prime} = \tilde{K}_1^{\prime \prime}
\tilde{A}_1^{\prime \prime} \tilde{K}_2^{\prime \prime}$ and
$\tilde{K}_2^{\prime} = \tilde{K}_3^{\prime \prime}
\tilde{A}_2^{\prime \prime} \tilde{K}_4^{\prime \prime}$. We finally
get the decomposition of $\tilde X_{sw}$, \be{finaldec} \tilde
X_{sw} = \tilde{K}_1^{\prime \prime} \tilde{A}_1^{\prime \prime}
\tilde{K}_2^{\prime \prime} \tilde{A}^{\prime} \tilde{K}_3^{\prime
\prime} \tilde{A}_2^{\prime \prime} \tilde{K}_4^{\prime \prime}, \ee
where
\begin{eqnarray} \label{Z11}
\tilde{A}_{1}^{\prime \prime} &=& diag \{ \begin{pmatrix}
    1 & 0 & 0 & 0 \\
    0 & 0 & 0 & 1 \\
    0 & 0 & 1 & 0 \\
    0 & -1 & 0 & 0
\end{pmatrix}, \begin{pmatrix}
    0 & 0 & -1 & 0 \\
    0 & 0 & 0 & 1 \\
    1 & 0 & 0 & 0 \\
    0 & -1 & 0 & 0
\end{pmatrix} \}, \nonumber \\
\tilde{A}_{2}^{\prime \prime} &=& diag \{ {\bf 1}_{4 },
\begin{pmatrix}
    1 & 0 & 0 & 0 \\
    0 & 0 & 0 & 1 \\
    0 & 0 & 1 & 0 \\
    0 & -1 & 0 & 0
\end{pmatrix} \}, \nonumber \\
\tilde{K}_{1}^{\prime \prime} &=& diag\{ {\bf 1_{2}},
\begin{pmatrix}
    0 & -1 \\
    1 & 0
\end{pmatrix}, \begin{pmatrix}
    0 & -1 \\
    1 & 0
\end{pmatrix}, \begin{pmatrix}
    0 & 1 \\
    -1 & 0
\end{pmatrix} \}, \\
\tilde{K}_{2}^{\prime \prime} &=& diag \{ {\bf 1}_{2},
\begin{pmatrix}
    0 & -1 \\
    1 & 0
\end{pmatrix}, {\bf 1}_{2}, {\bf 1}_{2 } \}, \nonumber \\
\tilde{K}_{3}^{\prime \prime} &=& diag \{ {\bf 1}_{2}, {\bf 1}_{2},
{\bf 1}_{2}, \begin{pmatrix}
    0 & -1 \\
    1 & 0
\end{pmatrix} \}, \nonumber \\
\tilde{K}_{4}^{\prime \prime} &=& diag \{ {\bf 1}_{2}, {\bf 1}_{2},
\begin{pmatrix}
    0 & -1 \\
    1 & 0
\end{pmatrix}, {\bf 1}_{2} \}. \nonumber
\end{eqnarray}
We now use the transformation $R$ in (\ref{cob}) to write $X_{sw}$
in the original coordinates as \be{almfin} X_{sw} = K_{1}^{\prime
\prime}A_{1}^{\prime \prime}K_{2}^{\prime \prime}A^{\prime}
K_{3}^{\prime \prime} A_{2}^{\prime \prime} K_{4}^{\prime \prime},
\ee with $A^{\prime} = R^T \tilde A^{\prime} R$, $A^{\prime\prime} =
R^T \tilde A^{\prime\prime} R$ and $K_{j}^{\prime \prime} = R^T
\tilde K_{j}^{\prime \prime} R$, $j = 1, 2$. We can write all the
factors in (\ref{almfin}) as exponentials of appropriate matrices in
the tensor product basis:
\begin{eqnarray}\label{aggiunta}
A^{\prime} = e^{a^{\prime}}, \quad A_k^{\prime \prime} =
e^{a_k^{\prime \prime}}, \quad k = 1, 2, \nonumber \\ K_j^{\prime
\prime} = e^{k^{\prime \prime}_j}, \quad j = 1, \ldots, 4
\end{eqnarray}
where
\begin{eqnarray}\label{finallist}
a^{\prime} &=& \frac{\pi}{2}(E_{11} \otimes \Delta_{24}) + \frac{3
\pi}{2} (E_{22} \otimes \Delta_{13}) + \pi(E_{22}\otimes
\Delta_{24}), \nonumber \\
a_1^{\prime \prime} &=& \frac{\pi}{4} (\Delta_{12} \otimes
\Omega_{24} + \Omega_{12} \otimes \Delta_{24}) + \frac{\pi}{4}
(\Delta_{12} \otimes \Omega_{24} - \Omega_{12} \otimes \Delta_{24})
+ \frac{3\pi}{4} (\Delta_{12} \otimes \Omega_{13} - \Omega_{12}
\otimes \Delta_{13}), \nonumber \\
a_2^{\prime \prime} &=& \frac{\pi}{4}(\Delta_{12}\otimes \Omega_{24}
- \Omega_{12}\otimes \Delta_{24}), \nonumber \\
k_1^{\prime \prime} &=& \frac{3 \pi}{2} (E_{22}
\otimes \Delta_{34}) + \frac{ \pi}{2} (E_{22} \otimes \Delta_{12}) +
\frac{3 \pi}{2}
(E_{11} \otimes \Delta_{34}), \\
k_2^{\prime \prime} &=& \frac{3\pi}{2} (E_{22} \otimes \Delta_{34}),
\nonumber \\
k_3^{\prime \prime} &=& \frac{3\pi}{2}(E_{22} \otimes \Delta_{12}), \nonumber \\
k_4^{\prime \prime} &=& \frac{3 \pi}{2}(E_{11} \otimes \Delta_{34}).
\nonumber
\end{eqnarray}

It is interesting to observe what number of nonlocal transformations
are needed to perform the given task if we are able to perform any
local transformation on the two subsystems. Notice that we are
considering the system as a bipartite system of a two level system
with a four level system. In essence, we assume that we have to
decide appropriate interactions between the two subsystems (two an
four dimensional) which along with local transformations will allow
us to perform the given task. By grouping the matrices that are
equivalent through local similarity transformations it is clear that
the Hamiltonians
\begin{eqnarray}\label{Hamil1}
H_1 &:=& E_{11} \otimes \Delta_{34}, \nonumber \\
H_2 &:=& \Delta_{12} \otimes \Omega_{24} + \Omega_{12} \otimes
\Delta_{24}, \\
H_3 &:=& \Delta_{12} \otimes \Omega_{24} - \Omega_{12} \otimes
\Delta_{24} \nonumber
\end{eqnarray}
are sufficient \footnote{In view of (\ref{finallist}) one can
replace $H_2$ with $\tilde H_{2} := \Delta_{12} \otimes
\Omega_{24}$}.




\begin{thebibliography}{99}

\bibitem{ConfraId} F. Albertini and D. D'Alessandro,
{\it Linear Algebra and its Applications}, Vol. 394, 237 (2005).

\bibitem{BBO1} S. Bullock and G. Brennen,
J. Math. Phys. 45, 2447 (2004).

\bibitem{MikoMCSS} D. D'Alessandro,
{\it Mathematics of Control, Signals and Systems}, (2003), 16, 1-25.

\bibitem{Mikofigata} D. D'Alessandro and F. Albertini,
xxx.lanl.gov, quant-ph/0504044,  submitted for
publication.

\bibitem{Helgason} S. Helgason, {\it Differential Geometry, Lie
Groups and Symmetric Spaces}, Academic Press, London, 1978.

\bibitem{KG} N. Khaneja and S. Glaser,
J. Chem. Phys. 267, 11 (2001).

\bibitem{Lloyd} S. Lloyd,
Phys. Rev. Lett. 75, 346 (1995).

\bibitem{RamaK} V. Ramakrishna, K. Flores, H. Rabitz and R. J.
Ober, Phys. Rev. A62, 053409 (2000).

\bibitem{Schirm} S. Schirmer, A.D. Greentree, V. Ramakrishna,
H. Rabitz, J. of Phys. A35, 8315 (2002).

\bibitem{Zhang} J. Zhang, J. Vala, S. Sastry, and K. B. Whaley
Phys. Rev. A67, 042313 (2003).

\end{thebibliography}
\end{document}